\documentclass{article}%
\usepackage[top=3cm, bottom=3cm, left=3cm, right=3cm]{geometry}
\usepackage{amsmath}
\usepackage{amsfonts}
\usepackage{amssymb}
\usepackage{graphicx}%
\begin{document}

\title{Phenomenology of axial-vector and pseudovector mesons: decays and mixing in
the kaonic sector}
\author{Florian Divotgey, Lisa Olbrich, and Francesco Giacosa\\\emph{Institut f\"{u}r Theoretische Physik, Johann Wolfgang Goethe -
Universit\"{a}t}\\\emph{Max von Laue--Str. 1, 60438 Frankfurt am Main, Germany}}
\maketitle

\begin{abstract}
We study the decays of the lightest axial-vector and pseudovector mesons,
interpreted as quark-antiquark states, into a vector and a pseudoscalar meson.
We show that the quarkonium assignment delivers a good description of the
decays and allows also to make further testable predictions. In the kaonic
sector, the physical resonances $K_{1}(1270)$ and $K_{1}(1400)$ emerge as
mixed objects of an axial vector state $K_{1,A}$ and a pseudovector state
$K_{1,B}$. We determine the mixing angle as $\left\vert \theta_{K}\right\vert
=\left(  33.6\pm4.3\right)  ^{\circ}$, a value compatible with previous
studies but with a smaller uncertainty. This result may be helpful for testing
models beyond the Standard Model of particle physics in which decays into the
kaonic resonances $K_{1}(1270)$ and $K_{1}(1400)$ are investigated.

\end{abstract}

\section{Introduction}

The understanding of hadron masses and decays using hadronic models, which
embody some of the symmetries of the underlying theory of quarks and gluons
(Quantum Chromodynamics, QCD), is an important task of modern particle
physics. The question about the nature of hadrons, if it is described in terms
of the `old quark model' or if new exotic hadrons, such as glueballs, hybrids,
but also tetraquark states are necessary, is in the center of a vivid
theoretical and (ongoing as well as planned) experimental efforts
\cite{amslerrev}. In addition, the emergence of dynamically generated
resonances due to meson interactions has attracted much theoretical attention:
bumps, which do not correspond to preformed quark substructure, can emerge due
to unitarity (loop) corrections on top of preexisting `seed states', see e. g.
Ref. \cite{klempt,dynrec,rupperice} and refs. therein.

In particular, $p$-wave quark-antiquark states represent an important subject
of hadron spectroscopy \cite{burakovsky}. The properties of $p$-wave
quark-antiquark states are rather different according to which nonet is taken
into consideration and at the two extremes there are the scalar and the tensor
mesons. Namely, on the one hand scalar $p$-wave quarkonia (total angular
momentum $L=1$ and total spin $S=1$ coupled to $J^{PC}=0^{++}$) offer
long-standing puzzles in hadron spectroscopy. They have been widely
investigated both with theoretical models, see for instance Refs.
\cite{amslerrev,dynrec,burakovsky,scalars,otherscalars,denis,dick}, and with
lattice QCD, e.g. Ref. \cite{latticescalar} and refs. therein. Being the
chiral partners of the pseudoscalar mesons, they are extremely important in
QCD also in relation to chiral symmetry breaking and its restoration at
nonzero temperature and density. Moreover, the emergence of further scalar
fields, such as the scalar glueball and scalar tetraquark states, has rendered
the whole scalar sector of QCD at the same time rich and difficult. On the
other hand, the lowest nonet of tensor mesons ($L=1,$ $S=1$ coupled to
$J^{PC}=2^{++}$) represents one of the best established ground state
quark-antiquark nonets with a very good agreement between the quark-antiquark
assignment and the measured masses and decay widths
\cite{burakovsky,quarkmodel,tensor,tensorlast}. (The situation is, however,
less clear when going to excited tensor states, see the discussion in Ref.
\cite{coitoacta}.)

In this work we concentrate our attention to the two remaining nonets of
$p$-wave states, the axial-vector and pseudovector quarkonia mesons. They are
not as enigmatic as the scalar ones but also not unambiguous as the tensor
states; in addition, interesting mixing effects between them are present. We
study the quarkonia axial-vector and pseudoscalar mesons by using a
phenomenological flavor symmetric relativistic Lagrangian in which the
axial-vector and the pseudovector mesons are introduced as standard
quark-antiquark fields. The Lagrangian is built in agreement with parity,
charge conjugation, flavor symmetry, and to the explicit breaking of the
latter due to different quark masses $(m_{s}>m_{u}=m_{d}$). In particular, we
shall study the three subjects that we present in the following.

(i) Phenomenology of the ground state axial-vector quark-antiquark fields:
these quarkonia mesons correspond to the quantum numbers $L=1$ and $S=1$
coupled to $J^{PC}=1^{++}$. They are the chiral partners of the vector mesons,
and are therefore part of chiral models in which the vector d.o.f. are
included, see Refs. \cite{geffen,ko} and the recently developed extended
Linear Sigma Model \cite{denis,dick} (in particular, in Ref. \cite{dick} for
the first time a linear chiral model with three flavors and (axial-)vector
quarkonia d.o.f. was studied). Still, a debate on the nature of the lightest
axial-vector states is ongoing: in our approach the isovector $(I=1)$
axial-vector state $a_{1}\equiv a_{1}(1260)$ is described by the
quark-antiquark isotriplet states $u\bar{d},$ $\bar{u}d,$ $\sqrt{1/2}(\bar
{u}u-\bar{d}d);$ however, the very same resonance has been interpreted as a
$\rho\pi$ quasi-bound state in the works of Ref. \cite{dyngena1}, see also the
general discussion of Ref. \cite{dynrec} about dynamically generated and/or
reconstructed states. In the present manuscript we test to which extent the
decays of $a_{1}$ are in agreement with the measured decay widths by utilizing
the quark-antiquark assignment in a relativistic approach. The very same
question can be extended to the two isoscalar members of the nonet,
$f_{1}(1285)$ and $f_{1}(1420)$, and to the axial-vector kaonic state
$K_{1,A}$ (present in both resonances $K_{1}(1270)$ and $K_{1}(1400)$, see
below). Our findings confirm the quark-antiquark nature of these axial-vector resonances.

(ii) Phenomenology of the ground state pseudovector quark-antiquark fields:
these quarkonia mesons correspond to the quantum numbers $L=1$ and $S=0$,
which implies that $J^{PC}=1^{+-}$. The quarkonium nature of the lightest
isovector states $b_{1}(1235)$, the isoscalar states $h_{1}(1170)$,
$h_{1}(1380)$ and the isodoublet state $K_{1,B}$ (present in both resonances
$K_{1}(1270)$ and $K_{1}(1400),$ see below) is established
\cite{burakovsky,quarkmodel,sukuki}. Our approach confirms these results: a
good description of the decay widths is obtained by simply using the
constraints of flavor symmetry in the channels in which experimental data are available.

(iii) Mixing in the kaonic sector: an interesting property which links the
two, otherwise separated, nonets of axial-vector and pseudovector mesons is
the fact that the two isodoublet axial-vector and pseudovector states
$K_{1,A}$ and $K_{1,B}$ mix and generate the two physical resonances
$K_{1}(1270)$ and $K_{1}(1400)$
\cite{burakovsky,quarkmodel,sukuki,cheng,hatanaka,chengyang,chengrevisiting}.
Within our model this mixing is generated by a flavor-symmetry breaking term,
which is proportional to $m_{s}^{2}-m_{u}^{2}.$ This also explains why the
mixing only takes place in the $I=1/2$ sector. Namely, a mixing between the
charged states $a_{1}^{\pm}$ and $b_{1}^{\pm}$ is proportional to $m_{d}%
^{2}-m_{u}^{2}$; it is then suppressed and neglected in this work. The neutral
states $a_{1}^{0},$ $b_{1}^{0},$ $f_{1}(1285)$ $f_{1}(1420)$ $h_{1}(1170)$,
$h_{1}(1380)$ are eigenstates of the charge conjugation operator $C$ and do
not mix, because $C$ is a conserved quantity in QCD.

We also calculate the mixing angle by studying the decays of $K_{1}(1270)$ and
$K_{1}(1400)$. In the notation of Refs.
\cite{burakovsky,sukuki,cheng,hatanaka,chengyang,chengrevisiting} usually
adopted in the literature, the values $\left\vert \theta_{K}\right\vert
\sim35^{\circ}$ and $\left\vert \theta_{K}\right\vert \sim55^{\circ}$ have
been obtained in a variety of phenomenological models, which have used the
masses and strong decays of the quarkonia nonets and also decays of $\tau$
lepton and, more recently, the decays of heavy $D$-type and $B$-type mesons.
In this work, a unique answer for the absolute value of the mixing angle is
obtained from all available strong decays: $\left\vert \theta_{K}\right\vert
=\left(  33.6\pm4.3\right)  ^{\circ}$. This result is in agreement with the
recent discussion of Ref. \cite{chengrevisiting}, in which the solution close
to $55^{\circ}$ has been shown to be not favoured, and also to the result of
Ref. \cite{hatanaka}, in which, using the weak decays of $B$ mesons, the value
and also the sign have been determined as $\theta_{K}=-(34\pm13)^{\circ}$.

Quite remarkably, the knowledge of this mixing angle is not only relevant for
a better understanding of hadron physics, but is also an important information
for testing the existence of new particles beyond the Standard Model and has
been for this reason subject of recent investigations
\cite{hatanakanew,ahmednew,lihuanew}. Namely, the transitions of heavy
$B$-mesons to axial- and pseudovector mesons can shed light to details of the
CKM\ matrix and to the presence of a fourth generation. In this view, our
approach represents an `up to date' and complete hadronic-based way to
determine the angle and this result can be of interest for studies of physics
beyond the Standard Model as well.

In order to achieve the goals (i)-(iii) described above we first calculate the
expression of the tree-level decay widths of axial- and pseudovector states
into a vector and a pseudoscalar meson as it follows from our interaction
Lagrangian; however, the use of simple tree-level expressions of the decay
widths is not enough for our purposes because many decays are affected by
threshold effects: the (nominal) masses of the produced particles are just
below (or even above) the threshold for their production. For this reason, an
integration over their spectral functions \cite{salam,achasov,lupo,duecan} is
necessary. To this end we use a relativistic version of the Breit-Wigner
distribution to which the corresponding energy threshold and the necessary
normalization are implemented.

The paper is organized as follows: in Sec. 2 we present the Lagrangian and the
employed decay formulae which take finite width effects into account; details
of the Lagrangian are presented in an Appendix. In Sec. 3 we present our
results for the decays of isoscalar and isovector states, and in Sec. 4 we
determine the mixing and the decays in the isodoublet (kaonic) sector.
Finally, in\ Sec. 5 we present our conclusions.

\section{The model: Lagrangian and decay widths}

The resonances we are interested in are the axial-vector and pseudovector
quark-antiquark nonets $A^{\mu}$ and $B^{\mu}$. These states decay
predominantly into a vector and a pseudoscalar meson, which are part of the
corresponding quark-antiquark nonets $V^{\mu}$ and $P.$ The four nonets are
described by the following matrices:
\begin{equation}
\begin{aligned} &P = \frac{1}{\sqrt{2}}\begin{pmatrix} \frac{\eta_{N} + \pi^{0}}{\sqrt{2}} & \pi^{+} & K^{+} \\ \pi^{-} & \frac{\eta_{N} - \pi^{0}}{\sqrt{2}} & K^{0} \\ K^{-} & \bar{K}^{0} & \eta_{S} \end{pmatrix}\text{ ,} &\quad &V^{\mu} = \frac{1}{\sqrt{2}}\begin{pmatrix} \frac{\omega^{\mu}_{N} + \rho^{\mu0}}{\sqrt{2}} & \rho^{\mu+} & K^{\mu\star+} \\ \rho^{\mu-} & \frac{\omega^{\mu}_{N} - \rho^{\mu0}}{\sqrt{2}} & K^{\mu\star0} \\ K^{\mu\star-} & \bar{K}^{\mu\star0} & \omega^{\mu}_{S} \end{pmatrix}\text{ ,} \\ &A^{\mu} = \frac{1}{\sqrt{2}}\begin{pmatrix} \frac{f^{\mu}_{1N,A} + a^{\mu0}_{1}}{\sqrt{2}} & a^{\mu+}_{1} & K^{\mu+}_{1,A} \\ a^{\mu-}_{1} & \frac{f^{\mu}_{1N,A} - a^{\mu0}_{1}}{\sqrt{2}} & K^{\mu0}_{1,A} \\ K^{\mu-}_{1,A} & \bar{K}^{\mu0}_{1,A} & f^{\mu}_{1S,A} \end{pmatrix}\text{ ,} &\quad &B^{\mu} = \frac{1}{\sqrt{2}}\begin{pmatrix} \frac{f^{\mu}_{1N,B} + b^{\mu0}_{1}}{\sqrt{2}} & b^{\mu+}_{1} & K^{\mu+}_{1,B} \\ b^{\mu-}_{1} & \frac{f^{\mu}_{1N,B} - b^{\mu0}_{1}}{\sqrt{2}} & K^{\mu0}_{1,B} \\ K^{\mu-}_{1,B} & \bar{K}^{\mu0}_{1,B} & f^{\mu}_{1S,B} \end{pmatrix}\text{ .} \end{aligned} \label{non}%
\end{equation}

The assignment of the fields is as follows: The matrix $P$ refers to the usual
nonet of pseudoscalar states $\{\pi,K,\eta,\eta^{\prime}\}$. The mixing of
strange and non-strange isoscalar sector implies that: $\eta=\eta_{N}%
\cos\varphi_{P}+\eta_{S}\sin\varphi_{P}$ and $\eta^{^{\prime}}=-\eta_{N}%
\sin\varphi_{P}+\eta_{S}\cos\varphi_{P}$, where $\eta_{N}$ is the pure
nonstrange state $\sqrt{1/2}(\bar{u}u+\bar{d}d)$ and $\eta_{S}$ the pure
strange state $\bar{s}s$. We use the numerical value $\varphi_{P}%
=-41.4^{\circ}$, which has been evaluated by the KLOE collaboration in\ Ref.
\cite{kloe}. Varying this mixing angle between the phenomenological range
$(-36^{\circ},-45^{\circ})$ generates only small numerical changes.

The matrix $V^{\mu}$ represents the vector states $\{\rho,K^{\ast}%
(892),\omega,\phi\}$, where $\omega$ is regarded as the purely nonstrange
state and $\phi$ as a purely strange state; we thus neglect the (small) mixing
angle in the vector sector. Finally, the matrix $A^{\mu}$ contains the nonet
of resonances of axial-vector states:
\[
\{a_{1}(1230),\text{ }K_{1,A},\text{ }f_{1}(1285),\text{ }f_{1}(1420)\}\text{
,}%
\]
and the matrix $B^{\mu}$ describes the nonet of pseudovector resonances
\[
\{b_{1}(1230),\text{ }K_{1,B},\text{ }h_{1}(1170),\text{ }h_{1}(1380)\}\text{
.}%
\]
In both nonets the strange-nonstrange isoscalar mixing is neglected, thus
$f_{1}(1285)$ and $h_{1}(1170)$ are purely nonstrange states, while
$f_{1}(1420)$ and $h_{1}(1380)$ are purely strange states.

The transformation properties of these nonets under charge, parity and flavor
transformations are summarized in Table 1.\begin{table}[h]
\makebox[\textwidth][c] {%
\begin{tabular}
[c]{|c|c|c|c|}\hline
& Parity (P) & Charge conjugation (C) & Flavor ($U\left(  3\right)  _{V}%
$)\\\hline\hline
$P$ & $-P(t,-\vec{x})$ & $P^{t}$ & $UPU^{\dagger}$\\\hline
$V^{\mu}$ & $V_{\mu}(t,-\vec{x})$ & $-\left(  V^{\mu}\right)  ^{t}$ &
$UV^{\mu}U^{\dagger}$\\\hline
$A^{\mu}$ & $-A_{\mu}(t,-\vec{x})$ & $\left(  A^{\mu}\right)  ^{t}$ &
$UA^{\mu}U^{\dagger}$\\\hline
$B^{\mu}$ & $-B_{\mu}(t,-\vec{x})$ & $-\left(  B^{\mu}\right)  ^{t}$ &
$UB^{\mu}U^{\dagger}$\\\hline
\end{tabular}
}\caption{Transformation properties of the nonets (\ref{non}) under charge,
parity and flavor transformations. The position of the indices is important
for parity: for instance, $V^{\mu}$ has odd-parity for $\mu=1,2,3$ but has
even-parity for $\mu=0.$}%
\end{table}

The Lagrangian describing the decay of axial- and pseudovector states into
vector and pseudoscalar mesons and the mixing in the kaonic sector consists of
three parts, each one containing one free parameter:
\begin{equation}
\mathcal{L}=\mathcal{L}_{A}+\mathcal{L}_{B}+\mathcal{L}_{mix}\text{ ,}
\label{lag}%
\end{equation}
where:

(i)%
\begin{equation}
\mathcal{L}_{A}=ia\;\mathrm{Tr}\left\{  A_{\mu}\left[  V^{\mu},P\right]
_{-}\right\}
\end{equation}
describes the coupling of the axial-vector fields to vector and pseudoscalar
ones; the unknown coupling constant is the parameter $a$ with the dimension of
energy; $\mathcal{L}_{A}$ is invariant under $P,$ $C,$ and $U(3)_{V}$. The
symbol $\left[  ,\right]  _{-}$ is the usual commutator. For the explicit form
of $\mathcal{L}_{A}$ see Appendix A.

(ii)
\begin{equation}
\mathcal{L}_{B}=b\;\mathrm{Tr}\left\{  B_{\mu}\left[  V^{\mu},P\right]
_{+}\right\}
\end{equation}
is the analogous Lagrangian, with coupling constant $b$, generating the
interaction of pseudovector fields with vector and pseudoscalar ones.
$\mathcal{L}_{B}$ is also invariant under $P,$ $C,$ and $U(3)_{V}.$ The symbol
$\left[  ,\right]  _{+}$ is the anticommutator. For the explicit form of
$\mathcal{L}_{B}$ see Appendix A.

(iii)%
\begin{equation}
\mathcal{L}_{mix}=ic\;\mathrm{Tr}\left\{  \Delta\;\left[  A_{\mu},B^{\mu
}\right]  _{-}\right\}  \text{ ,} \label{mixlag}%
\end{equation}
in which $c$ is a dimensionless coupling constant and $\Delta$ is a diagonal
matrix with the bare quark masses $\Delta=\mathrm{diag\{}m_{u}^{2},m_{d}%
^{2},m_{s}^{2}\}$. $\mathcal{L}_{mix}$ is still invariant under $P$ and $C$
transformations, but breaks the symmetry under $U(3)_{V}$ transformations when
the bare quark masses are not equal. Notice that, in the limit in which all
the quark masses coincide, the mixing Lagrangian vanishes. Here, we work in
the isospin symmetric limit $m_{u}=m_{d}.$ In this limit the Lagrangian takes
the form
\begin{equation}
\mathcal{L}_{mix}=\frac{ic}{2}\left(  m_{s}^{2}-m_{u}^{2}\right)  \left\{
K_{1,A\mu}^{-}K_{1,B}^{\mu+}-K_{1,A\mu}^{+}K_{1,B}^{\mu-}+\bar{K}_{1,A\mu}%
^{0}K_{1,B}^{\mu0}-K_{1,A\mu}^{0}\bar{K}_{1,B}^{\mu0}\right\}  \label{mixlag2}%
\end{equation}
As a consequence, only the kaonic fields $K_{1,A}$ and $K_{1,B}$ mix and
generate the two physical resonances $K_{1}(1270)$ and $K_{1}(1400)$:%
\begin{equation}%
\begin{pmatrix}
K_{1}^{\mu}(1270)\\
K_{1}^{\mu}(1400)
\end{pmatrix}
=%
\begin{pmatrix}
\cos\varphi & -i\sin\varphi\\
-i\sin\varphi & \cos\varphi
\end{pmatrix}%
\begin{pmatrix}
K_{1,A}^{\mu}\\
K_{1,B}^{\mu}%
\end{pmatrix}
\text{ .} \label{mixrel1}%
\end{equation}
The link of the here employed mixing angle $\varphi$ to the mixing angle
$\theta_{K}$, introduced in the literature
\cite{burakovsky,sukuki,cheng,hatanaka,chengyang,chengrevisiting} as
\begin{equation}%
\begin{pmatrix}
\left\vert K_{1}^{+}(1270)\right\rangle \\
\left\vert K_{1}^{+}(1400)\right\rangle
\end{pmatrix}
=%
\begin{pmatrix}
\sin\theta_{K} & \cos\theta_{K}\\
\cos\theta_{K} & -\sin\theta_{K}%
\end{pmatrix}%
\begin{pmatrix}
\left\vert K_{1,A}^{+}\right\rangle \\
\left\vert K_{1,B}^{+}\right\rangle
\end{pmatrix}
\text{ ,} \label{thetak0}%
\end{equation}
is given by%
\begin{equation}
\theta_{K}=90^{\circ}+\varphi\text{ .} \label{thetak}%
\end{equation}

In the context of mixing a comment is important: Our mixing Lagrangian
describes this mixing at tree-level. However, a mixing between $K_{1,A}$ and
$K_{1,B}$ arises when loops are taken into account: namely, as we shall
describe later on, both states $K_{1,A}$ and $K_{1,B}$ couple predominantly to
the same final state $K^{\ast}\pi$ and thus a loop of $K^{\ast}$ and $\pi$ can
transform $K_{1,A}$ into $K_{1,B}$ (and vice-versa). (Such kind of mixing is a
quantum mixing; it appears also in other contexts, see for instance Ref.
\cite{hanhart} for the $a0(980)$-$f_{0}(980)$ system, e.g. Ref. \cite{k0} for
the neutral kaon system, and Ref. \cite{rupp} for axial- and pseudovector
$D_{s}$ states). In this work we do not evaluate this loop and we effectively
describe this mixing by the constant term in Eq. (\ref{mixlag}). Going beyond
this approximation would be an interesting task for the future.

We now turn to the decays of a resonance $R$ belonging to the axial-vector
nonet $A^{\mu}$ or to pseudovector nonet $B^{\mu}$. The tree-level decay width
of the resonance $R$ decaying into a vector ($V$) and pseudoscalar ($P$)
particle is given by the formula
\begin{equation}
\Gamma_{R\rightarrow VP}^{\text{tl}}(m_{R},m_{V},m_{P})=\frac{g_{RVP}^{2}%
k_{f}(m_{R},m_{V},m_{P})}{24\pi m_{R}^{2}}\left[  2+\frac{\left(  m_{R}%
^{2}+m_{V}^{2}-m_{P}^{2}\right)  ^{2}}{4m_{R}^{2}m_{V}^{2}}\right]
\Theta\left(  m_{R}-m_{V}-m_{P}\right)  \text{ ,} \label{widthallg}%
\end{equation}
where $m_{R},m_{V},m_{P}$ refer to the masses of the initial ($R)$ and final
($P$ and $V$) states of the decay process and $\Theta\left(  x\right)  $ is
the step function; the coupling constant $g_{RVP}$ is a function of the
constants $a,$ $b$ and $c$ entering in Eq. (\ref{lag}) and its explicit value
can be read off, case by case, in the full expression in Appendix A (by
convention, isospin degeneracy factors are also formally included in $g_{RVP}%
$). Finally, the momentum function $k_{f}(m_{R},m_{V},m_{P})$ is the modulus
of the three-momentum of an outgoing particle:
\begin{equation}
k_{f}(m_{R},m_{V},m_{P})=\frac{1}{2m_{R}}\sqrt{m_{R}^{4}+\left(  m_{V}%
^{2}-m_{P}^{2}\right)  ^{2}-2m_{R}^{2}\left(  m_{V}^{2}+m_{P}^{2}\right)
}\text{ .} \label{tldw}%
\end{equation}

The use of the tree-level decay width is satisfactory when threshold effects
are negligible. Being this not the case for some of the decay channels studied
in this work, we include the finite width effects by integrating the decay
width weighted with the spectral functions of the involved (axial-)vector and
pseudovector particles over the particle masses \cite{lupo}. Pseudoscalar
mesons are taken as stable in view of their very small decay width. For what
concerns the form of the spectral function of the other fields, we use a
relativistic Breit-Wigner form with the proper corresponding energy
threshold:
\begin{equation}
\begin{aligned} d_{R}\left(x\right) &= \frac{N_{R}}{\left(x^{2} - m^{2}_{R}\right)^{2} + \left(m_{R} \Gamma_{R}\right)^2}\Theta\left(x - m_{threshold,R}\right) \text{ ,}\\ d_{V}\left(x\right) &= \frac{N_{V}}{\left(x^{2} - m^{2}_{V}\right)^{2} + \left(m_{V} \Gamma_{V}\right)^2}\Theta\left(x - m_{threshold,V}\right) \text{ .} \end{aligned} \label{specfunc1}%
\end{equation}

The constants $N_{R}$ and $N_{V}$ are chosen in a way so that $\int
_{0}^{\infty}{\mathrm{d}x\;d_{R/V}\left(  x\right)  }=1$. Finally, the decay
width reads
\begin{equation}
\Gamma_{R\rightarrow VP}=\int\limits_{0}^{\infty}\int\limits_{0}^{\infty
}\mathrm{d}x\mathrm{d}y\,\,\Gamma_{R\rightarrow VP}^{\text{tl}}\left(
x,y,m_{P}\right)  d_{R}\left(  x\right)  d_{V}\left(  y\right)  \text{ .}
\label{totalwidth}%
\end{equation}

Notice that in this work we do not take into consideration loop corrections to
the masses and decay widths. For what concerns the masses we shall use the
experimental values quoted by the PDG \cite{PDG} as inputs and do not attempt
a theoretical derivation, while for what concerns the decay widths the formula
(\ref{totalwidth}) represents a valid phenomenological description, as long as
the ratio of the decay width and the mass of the unstable state is not too
large, see Ref. \cite{wolkanowski} for a comparison of this treatment with the
position (in particular, the imaginary part) of the pole.

\section{Results for $I=0$ and $I=1$ resonances}

We now turn to the determination of the parameters and to the results of the
decays for isoscalar and isovector axial- and pseudovector states.

We first concentrate on axial-vector mesons as described by the Lagrangian in
Eq. (\ref{lag}): we study the decays of $f_{1}\left(  1420\right)  $,
$f_{1}\left(  1285\right)  $, and $a_{1}\left(  1260\right)  $. First of all,
we use the decay $f_{1}\left(  1420\right)  \rightarrow K\bar{K}^{\star
}\left(  892\right)  $ to determine the axial-vector coupling constant $a$.
The decay width for this channel is given by Eq. (\ref{tldw}) and
(\ref{specfunc1}) by setting $g_{RVP}=a/\sqrt{2}$ and by using the
corresponding PDG \cite{PDG} masses and decay widths (for the vector and
pseudoscalar kaons the masses of the charged kaon states are used). The
average of the total decay width of $f_{1}\left(  1420\right)  $ reported by
\cite{PDG} reads $\Gamma_{f_{1}\left(  1420\right)  }=\left(  54.9\pm
2.6\right)  $ MeV. The dominant decay channels of $f_{1}\left(  1420\right)  $
are given by the decay channels into $K\bar{K}^{\star}\left(  892\right)  $
and $K\bar{K}\pi$. Using the branching ratio between those two channels, we
derive the experimental value
\begin{equation}
\Gamma_{f_{1}\left(  1420\right)  \rightarrow K\bar{K}^{\star}\left(
892\right)  +c.c.}^{\exp}=\left(  44.5\pm4.2\right)  \text{ MeV,}
\label{width1}%
\end{equation}
out of which the coupling constant $a$ is determined:
\begin{equation}
\left\vert a\right\vert =\left(  5.43\pm0.26\right)  \text{ GeV.} \label{a}%
\end{equation}
In the next step we use this value for $a$ to determine the other decay widths
of axial-vector mesons and compare them to available experimental data. These
results for the $I=0$ and $I=1$ decays are summarized in Table 2.

\begin{table}[h]
\makebox[\textwidth][c] {%
\begin{tabular}
[c]{|c|c|c|}\hline
Decay process & Theory (MeV) & Experiment (MeV)\\\hline\hline
$f_{1}\left(  1420\right)  \rightarrow K\bar{K}^{\star}\left(  892\right)
+c.c$ & $44.5\pm4.2$ & $44.5\pm4.2$\\\hline
$a_{1}\left(  1260\right)  \rightarrow\rho\pi$ & $396\pm37$ & Dominant;
$\Gamma_{a_{1}}^{\text{tot}}=%
\begin{array}
[c]{c}%
250\text{-}600\text{ (estimate \cite{PDG})}\\
367\pm9_{-25}^{+28}\text{ (Ref. \cite{alekseev})}%
\end{array}
$\\\hline
$a_{1}\left(  1260\right)  \rightarrow K\bar{K}^{\star}\left(  892\right)
+c.c.$ & $32.1\pm3.03$ & $6$-$55$\\\hline
$f_{1}\left(  1285\right)  \rightarrow K\bar{K}^{\star}\left(  892\right)
+c.c$ & $2.79\pm0.26$ & not seen\\\hline
$f_{1}(1285)\rightarrow\rho\pi$ & $0$ & $<0.075$\\\hline
\end{tabular}
}\caption{Decays of $I=1$ and $I=0$ axial-vector states. The first
experimental entry has been used to fix the coupling constant $a.$}%
\end{table}

The following comments are in order:

a.1) The resonance $a_{1}(1260)$ is very broad and the dominant channel, both
experimentally and theoretically, is the $\rho\pi$ one. Our result for this
channel confirms this expectation and fits well in the estimated range of the
PDG \cite{PDG} and is in very good agreement with the recent experimental
result of the Compass Collaboration in Ref. \cite{alekseev}.

a.2) We also have evaluated the decay of $a_{1}(1260)$ into $K\bar{K}^{\ast
}(892)$. This decay mode has been seen; although no average or fit is
performed by the PDG, the listed experimental values obtained by various
experiments lie in the range between $5$ and $55$ MeV \cite{PDG}. Our
theoretical result is in agreement with these results. Notice that this decay
mode could not be calculated without using the spectral functions because the
sum of the nominal masses of the final state is higher than the nominal mass
of the decaying particle.

a.3) In our theoretical framework the resonance $f_{1}\left(  1285\right)  $
couples only to the channel $K\bar{K}^{\ast}(892).$ We thus evaluated the
decay width, which is about $3$ MeV. Experimentally, this decay has not yet
been seen \cite{PDG}. Note, the full decay width of $f_{1}\left(  1285\right)
$ amounts to $(24.2\pm1.1)$ MeV, thus very narrow. The fact that this
resonance is narrow is well explained in our quarkonium framework: the
would-be dominant $K\bar{K}^{\ast}(892)$ channel is in this case very small
because it is kinematically suppressed. The dominant decay channels of
$f_{1}\left(  1285\right)  $ are the $4\pi$ and the $\eta\pi\pi$ ones, which
are not included in our model. The role of light scalar mesons is relevant for
these decays; for instance, for the $\eta\pi\pi$ channel the contribution of
$a_{0}(980)$ is sizable \cite{PDG}.

a.4) It is interesting to stress that within our model $f_{1}\left(
1285\right)  $ does not couple to $\rho\pi$: this is a consequence of the
symmetries of our approach. Our theoretical prediction is thus zero. This is
very well verified by the experiment, for which a small upper limit has been set.

\bigskip

We now turn to the pseudovector sector. To fix the coupling constant $b$ we
use the total decay width of $b_{1}(1235)$: $\Gamma_{b_{1}(1235)}^{\exp
}=(142\pm9)$ MeV \cite{PDG}. The relevant channels contributing to the total
decay width are $\omega\pi$, $\rho\eta$ and $KK^{\star}(892)$. Thus, out of
the equation
\begin{equation}
\Gamma_{b_{1}(1235)\rightarrow\omega\pi}+\Gamma_{b_{1}(1235)\rightarrow
\eta\rho}+\Gamma_{b_{1}(1235)\rightarrow KK^{\star}(892)+c.c.}\overset{!}%
{=}(142\pm9)\,\text{MeV ,}%
\end{equation}
we obtain%
\begin{equation}
\left\vert b\right\vert =\left(  7.0\pm0.22\right)  \,\text{GeV .} \label{b}%
\end{equation}

We use this value of $b$ to determine the other decays of $I=0$ and $I=1$
pseudovector mesons. The results are summarized in Table \ref{pseudovecwidth}.
\begin{table}[h]
\makebox[\textwidth][c] {%
\begin{tabular}
[c]{|c|c|c|}\hline
Decay process & Theory (MeV) & Experiment (MeV)\\\hline\hline
$b_{1}(1235)\rightarrow\omega\pi$ & $110.0\pm7.0$ & dominant ($\Gamma
_{b_{1}(1235)}^{tot}=142\pm9$)\\\hline
$b_{1}(1235)\rightarrow\eta\rho$ & $18.4\pm1.2$ & seen\\\hline
$b_{1}(1235)\rightarrow K\bar{K}^{\ast}(892)+c.c.$ & $16.2\pm1.0$ &
seen\\\hline
$b_{1}(1235)\rightarrow\eta^{\prime}\rho$ & $1.07\pm0.07$ & not seen\\\hline
$b_{1}(1235)\rightarrow\phi\pi$ & $0$ & $<2.1$\\\hline
$h_{1}(1170)\rightarrow\rho\pi$ & $364\pm23$ & dominant ($\Gamma_{h_{1}%
(1170)}^{tot}=360\pm40)$\\\hline
$h_{1}(1170)\rightarrow\eta\omega$ & $7.15\pm0.46$ & not seen\\\hline
$h_{1}(1170)\rightarrow K\bar{K}^{\ast}(892)+c.c$ & $7.99\pm0.51$ & not
seen\\\hline
$h_{1}(1380)\rightarrow K\bar{K}^{\ast}(892)+c.c.$ & $54.0\pm3.4$ & $91\pm
30$\\\hline
$h_{1}(1380)\rightarrow\phi\eta$ & $2.28\pm0.13$ & not seen\\\hline
$h_{1}(1380)\rightarrow\phi\eta^{\prime}$ & $0.93\pm0.06$ & not seen\\\hline
\end{tabular}
}\label{pseudovecwidth}\caption{Decays of $I=1$ and $I=0$ pseudovector states.
The sum of the first three exerimental entries has been used to fix the
parameter $b.$}%
\end{table}

The following comments are in order:

b.1) The decay $b_{1}(1235)\longrightarrow\omega\pi$ is, according to PDG, the
dominant one. This is in well agreement with our result. In addition, we also
have predicted the decays of the processes $b_{1}(1235)\longrightarrow\eta
\rho$ and $b_{1}(1235)\rightarrow K\bar{K}^{\ast}(892)$, both of which have
been seen, but no branching ratio is reported in \cite{PDG}.

b.2) The theoretical ratio $\Gamma_{b_{1}(1235)\rightarrow\eta\rho}%
/\Gamma_{b_{1}(1235)\rightarrow\omega\pi}=0.17\pm0.02$ should be compared to
the experimental ratio $\Gamma_{b_{1}(1235)\rightarrow\eta\rho}/\Gamma
_{b_{1}(1235)\rightarrow\omega\pi}<0.10$ reported in \cite{PDG}. Our
theoretical value is close but slightly above this upper limit. It should be
however noticed that only one experiment has reported this ratio
\cite{atkinson}.

b.3) We also predict the ratio $\Gamma_{b_{1}(1235)\rightarrow KK^{\star
}(892)+c.c.}/\Gamma_{b_{1}(1235)\rightarrow\omega\pi}=0.15\pm2$, which has not
been measured yet.

b.4) The theoretical prediction for the channel $b_{1}(1235)\rightarrow
\eta^{\prime}\rho$ has been reported in Table 3. This channel is subleading
and has not yet been seen.

b.5) The decay channel $b_{1}(1235)\longrightarrow\phi\pi$ vanishes exactly in
our approach and is thus in agreement with the experimental upper limit.
Moreover, the very small experimental value for the ratio $\Gamma
_{b_{1}(1235)\rightarrow\phi\pi}/\Gamma_{b_{1}(1235)\rightarrow\omega\pi
}<0.004$ confirms the absence of this decay channel, in well agreement with
the quarkonium assignment.

b.6) The experimental decay $\Gamma_{h_{1}(1170)}^{exp}=(360\pm40)$ MeV is in
very well agreement with our theoretical result of $(373\pm24)$ MeV. Moreover,
only the channel $h_{1}(1170)\longrightarrow\rho\pi$ has been experimentally
seen: this is also reproduced by our results, where the decay modes
$h_{1}(1170)\longrightarrow\eta\omega$ and $h_{1}(1170)\longrightarrow
K\bar{K}^{\ast}(892)$ are about $7$ MeV (and thus much smaller than
$h_{1}(1170)\longrightarrow\rho\pi).$ Still, these decay modes are testable
predictions of our approach.

b.7) The resonance $h_{1}(1380)$ decays predominantly into $K\bar{K}^{\star
}(892)$; experiment and theory are in well agreement. In addition, we have
predicted the decay of the channels $h_{1}(1380)\rightarrow\phi\eta$ and
$h_{1}(1380)\rightarrow\phi\eta^{\prime}$, which turn out to be subdominant.

\bigskip

Summarizing, for both nonets the theoretical results are in good agreement
with the available measured ones. In addition, we could make predictions for
branching ratios of not-yet observed channels, which can be tested in future experiments.

\section{Mixing and results in the kaonic sector}

In this section we determine the mixing angle $\varphi$ defined in Eq.
(\ref{mixrel1}) in the kaonic sector and the corresponding decays of
$K_{1}\left(  1270\right)  $ and $K_{1}\left(  1400\right)  $.

As mentioned in Sec. 2, the non-vanishing difference of the bare quark masses
$m_{s}^{2}-m_{u}^{2}$ induces a mixing of the kaonic fields $K_{1,A}$ and
$K_{1,B}$. We perform a unitary transformation from the unphysical basis
$\left\{  K_{1,A},K_{1,B}\right\}  $ to the physical isodoublets $\left\{
K_{1}\left(  1270\right)  ,K_{1}\left(  1400\right)  \right\}  $ by using the
(inverse form) of Eq. (\ref{mixrel1}). For the determination of $\varphi$ we
use the $K^{\star}\left(  892\right)  \pi$ decay mode of the resonances
$K_{1}\left(  1270\right)  $ and $K_{1}\left(  1400\right)  $. The partial
decay widths are given by $\Gamma_{K_{1}\left(  1270\right)  \rightarrow
K^{\star}\pi}^{exp}=\left(  14.4\pm5.5\right)  $ MeV \cite{PDG} and
$\Gamma_{K_{1}\left(  1400\right)  \rightarrow K^{\star}\pi}^{exp}=\left(
117\pm10\right)  $ MeV \cite{carnegie}.

We perform a fit by minimizing the $\chi^{2}$-function,
\begin{equation}
\chi^{2}\left(  \varphi\right)  =\left(  \frac{\Gamma_{K_{1}\left(
1270\right)  \rightarrow K^{\star}\pi}^{th}\left(  \varphi\right)
-\Gamma_{K_{1}\left(  1270\right)  \rightarrow K^{\star}\pi}^{exp}}%
{\delta\Gamma_{K_{1}\left(  1270\right)  \rightarrow K^{\star}\pi}^{exp}%
}\right)  ^{2}+\left(  \frac{\Gamma_{K_{1}\left(  1400\right)  \rightarrow
K^{\star}\pi}^{th}\left(  \varphi\right)  -\Gamma_{K_{1}\left(  1400\right)
\rightarrow K^{\star}\pi}^{exp}}{\delta\Gamma_{K_{1}\left(  1400\right)
\rightarrow K^{\star}\pi}^{exp}}\right)  ^{2}\text{ ,} \label{chi}%
\end{equation}
which implies (restricting to the interval $[-\pi/2,\pi/2]$):
\begin{equation}
\left\vert \varphi_{1}\right\vert =\left(  56.4\pm4.2\right)  ^{\circ}%
\qquad,\qquad\left\vert \varphi_{2}\right\vert =\left(  19.0\pm4.2\right)
^{\circ}\text{ ,} \label{angle}%
\end{equation}
with the acceptable $\chi^{2}$-value $\chi^{2}\left(  \varphi_{1}\right)
=\chi^{2}\left(  \varphi_{2}\right)  =1.54$.

Notice that we cannot determine the sign of the mixing angle $\varphi$ because
we do not know the sign of $a$ and $b.$ In particular, $\varphi<0$ implies
$a\cdot b>0$ and $\varphi>0$ implies $a\cdot b<0.$

For both angles in\ Eq. (\ref{angle}) we calculate the decay widths of both
resonances $K_{1}\left(  1270\right)  $ and $K_{1}\left(  1400\right)  $ into
$K\rho$ and $K\omega.$ The results are presented in Table 4. Note, while for
the resonance $K_{1}\left(  1270\right)  $ we use the branching ratios of the
summarizing table and the estimated full width $\Gamma_{K_{1}\left(
1270\right)  }^{exp}=(90\pm20)$ MeV reported in Ref. \cite{PDG}, for the
resonance $K_{1}\left(  1400\right)  $ we use the results of Ref.
\cite{carnegie}, in which the partial decay widths of this resonance have been
directly determined, see also \cite{PDG}. \begin{table}[h]
\makebox[\textwidth][c] {%
\begin{tabular}
[c]{|c|c|c|c|}\hline
Decay process & \multicolumn{2}{c|}{Theory (MeV)} & Experiment
(MeV)\\\hline\hline
& $\left\vert \varphi_{1}\right\vert =\left(  56.4\pm4.3\right)  ^{\circ}$ &
$\left\vert \varphi_{2}\right\vert =\left(  19.5\pm4.3\right)  ^{\circ}$ &
\\\hline\hline
$K_{1}(1270)\rightarrow K^{\ast}\pi$ & $10.8\pm0.79$ & $10.8\pm0.79$ &
$14.4\pm5.5$ \cite{PDG}\\\hline
$K_{1}(1400)\rightarrow K^{\ast}\pi$ & $106.4\pm7.8$ & $106.4\pm7.8$ &
$117\pm10$ \cite{carnegie}\\\hline
$K_{1}(1270)\rightarrow K\rho$ & $56.2\pm4.1$ & $39.6\pm2.9$ & $38\pm10$
\cite{PDG}\\\hline
$K_{1}(1270)\rightarrow K\omega$ & $9.5\pm0.7$ & $6.74\pm0.49$ & $9.9\pm2.8$
\cite{PDG}\\\hline
$K_{1}(1400)\rightarrow K\rho$ & $0.45\pm0.03$ & $27.3\pm1.2$ & $2\pm1$
\cite{carnegie}\\\hline
$K_{1}(1400)\rightarrow K\omega$ & $0.13\pm0.01$ & $7.59\pm0.56$ & $23\pm12$
\cite{carnegie}\\\hline
\end{tabular}
}\caption{$K_{1}\left(  1270\right)  $ and $K_{1}\left(  1400\right)  $ decay
widths for $K\rho$ and $K\omega$ decay channel. The first two entries have
been used to determine the mixing angle $\varphi$ via a fit.}%
\end{table}

Following comments are in order:

(i) The case $\left\vert \varphi_{1}\right\vert =\left(  56.4\pm4.3\right)
^{\circ}$ (in turn: $\left\vert \theta_{K}\right\vert =\left(  33.6\pm
4.3\right)  ^{\circ}$). This large mixing angle implies that $K_{1}(1270)$ is
described by the pseudovector state $K_{1,B}$ with a probability of $69\%$ and
by the axial-vector state $K_{1,A}$ with the remaining probability of $29\%;$
a reversed situation holds for $K_{1}(1400).$ The masses of the bare states
$K_{1,A}$ and $K_{1,B}$ read%
\begin{equation}
m_{K_{1,A}}=1.36\text{ GeV , }m_{K_{1,B}}=1.31\text{ GeV ,}%
\end{equation}
thus realizing the bare ordering $m_{K_{1,B}}<m_{K_{1,A}}$. This bare level
ordering is also in agreement with the other members of the nonets:
$h_{1}(1170)$ and $h_{1}(1380)$ are lighter than the corresponding states
$f_{1}(1285)$ and $f_{1}(1420)$. The corresponding value of the parameter $c$
defined in Eq. (\ref{mixlag}) and (\ref{mixlag2}) reads
\begin{equation}
c=\frac{m_{K_{1,A}}^{2}-m_{K_{1,B}}^{2}}{m_{s}^{2}-m_{u}^{2}}\tan(2\varphi
_{1})=\pm\left(  35_{-11}^{+23}\right)  \text{ ,}%
\end{equation}
where for the quark masses we have used $m_{s}=95$ MeV and $m_{u}=2.3$ MeV and
where a positive sign of $c$ corresponds to a negative sign of $\varphi_{1}$
and vice-versa.

In Table 4 we have presented the theoretical results of all decay channels,
and the first five entries are in good agreement with the experiment. For the
last entry $K_{1}(1400)\rightarrow K\omega$ the theoretical prediction is
close to zero in virtue of a destructive interference; this result is not in
good agreement with the experimental result of Ref. \cite{carnegie}, being off
of a factor $2\sigma$. However, it should be stressed that the branching ratio
reported in the summarizing table of Ref. \cite{PDG} reads $\left(
1.0\pm1.0\right)  \%$, which is well compatible with our result.

(ii) The case $\left\vert \varphi_{2}\right\vert =\left(  19.5\pm4.5\right)
^{\circ}$ (in turn: $\left\vert \theta_{K}\right\vert =\left(  70.5\pm
4.3\right)  ^{\circ}$). This small mixing angle implies that $K_{1}(1270)$ is
described by the axial-vector state $K_{1,A}$ with a probability of $89\%$ and
by the pseudovector state $K_{1,B}$ with the remaining probability of $11\%;$
a reversed situation holds for $K_{1}(1400)$. The bare masses read%
\begin{equation}
m_{K_{1,A}}=1.29\text{ GeV , }m_{K_{1,B}}=1.39\text{ GeV .}%
\end{equation}
(The parameter $c$ reads now $c=\pm\left(  23_{-6}^{+11}\right)  $, where a
positive sing of $c$ corresponds to a negative sign of $\varphi_{2}$ and
vice-versa) However, there are clear reasons why this mixing angle is not
favoured: first of all, the decay mode $K_{1}(1400)\rightarrow K\rho$ reads
$27.3\pm1.2$ MeV, which is much higher than the experimental value $(2\pm1)$
MeV. Even using the branching ratio $0$-$6\%$ reported by the PDG, one finds
the upper limit of $10.5$ MeV for this decay channel. In addition, the bare
level ordering $m_{K_{1,A}}>$ $m_{K_{1,B}}$ is not expected. In the end, in
the literature a solution of the type $\left\vert \theta_{K}\right\vert
\sim70^{\circ}$ has not been found (typically, the alternative solution found
in many works reads, as discussed in the introduction, $\left\vert \theta
_{K}\right\vert \sim55^{\circ}$).

In conclusion, a overall good agreement is obtained for the $\left\vert
\varphi_{1}\right\vert =\left(  56.3\pm4.2\right)  ^{\circ}$, which means
$\left\vert \theta_{K}\right\vert =\left(  33.6\pm4.3\right)  ^{\circ}$. Using
the sign determination of Ref. \cite{hatanaka}, $\theta_{K}=-(34\pm13)^{\circ
}$, we are led to choose
\begin{equation}
\theta_{K}=\left(  -33.6\pm4.3\right)  ^{\circ}\text{ .}%
\end{equation}
In turn, this result means that the coupling constants $a$ and $b$ have
opposite signs: $a\cdot b<0$.

\section{Conclusions}

In this work we have studied the decays of axial-vector and pseudoscalar
mesons interpreted, as quark-antiquark states, into a vector and a
pseudoscalar meson. To this end, we have used a relativistic quantum field
theoretical model which makes use of flavor symmetry and its explicit breaking
due to non equal bare quark masses.

Our model contains three unknown parameters. Two of them describe the
interaction of axial-vector and pseudovector mesons with pseudoscalar and
vector states respectively. The results have been summarized in Table 2 and
Table 3, where a good agreement with the experiment is shown and further
theoretical predictions for not-yet measured decay channels have been
presented. The remaining free parameter describes the mixing of a bare
axial-vector kaonic state $K_{1,A}$ and a bare pseudovector kaonic state
$K_{1,B},$ leading to the resonances $K_{1}(1270)$ and $K_{1}(1400)$. The
results are summarized in Table 4, where it is shown that a good agreement in
all decay channels is achieved for the mixing angle $\left\vert \theta
_{K}\right\vert =\left(  33.6\pm4.3\right)  ^{\circ}.$

In conclusion, our results confirm the predominant quark-antiquark nature of
the ground-state axial-vector and pseudovector mesons. In the kaonic sector we
have provided an independent determination of the mixing angle, which is
potentially useful in the search and/or falsification of physics beyond the
Standard Model.

\bigskip

\textbf{Acknowledgment: }The authors thank Denis Parganlija and Dirk H.
Rischke for very useful discussions. F.G. thanks the support from the
Foundation Polytechnic Society of Frankfurt am Main through an Educator fellowship.

\newpage

\section*{Appendix A}

According to (\ref{lag}) and the discussion of section 2 our Lagrangian
consists of three parts:
\begin{equation}
\mathcal{L}=\mathcal{L}_{A}+\mathcal{L}_{B}+\mathcal{L}_{mix},\nonumber
\end{equation}
in which the axial-vector part of the Lagrangian is given by
\begin{equation}
\begin{aligned} \mathcal{L}_{A} = \; &ia\;\mathrm{Tr}\left\{ A_{\mu}\left[ V^{\mu},P\right]_{-}\right\} \\ = \; &\frac{ai}{2\sqrt{2}} \left\{ \frac{1}{\sqrt{2}} f_{1N,A\mu}\left(K^{\star\mu+}K^{-} - K^{\star\mu-}K^{+} + K^{\star\mu0}\bar{K}^{0} - \bar{K}^{\star\mu0}K^{0}\right) \right. + f_{1S,A\mu}\left(K^{\star\mu-}K^{+} - K^{\star\mu+}K^{-} \right. \\ &+ \left.\bar{K}^{\star\mu0}K^{0} - K^{\star\mu0}\bar{K}^{0}\right) + \frac{1}{\sqrt{2}}a^{0}_{1\mu}\left(2\rho^{\mu+}\pi^{-} - 2\rho^{\mu-}\pi^{+} + K^{\star\mu+}K^{-} - K^{\star\mu-}K^{+} - K^{\star\mu0}\bar{K}^{0} - \bar{K}^{\star\mu0}K^{0}\right) \\ &+ a^{+}_{1\mu}\left(\sqrt{2}\rho^{\mu-}\pi^{0} - \sqrt{2}\rho^{\mu0}\pi^{-} + K^{\star\mu0}K^{-} - K^{\star\mu-}K^{0}\right) + a^{-}_{1\mu}\left(\right.\sqrt{2}\rho^{\mu0}\pi^{+} - \sqrt{2}\rho^{\mu+}\pi^{0} + K^{\star\mu+}\bar{K}^{0} \\ &- \left.\bar{K}^{\star\mu0}K^{+}\right) + K^{+}_{1,A\mu}\left(\omega^{\mu}_{S}K^{-} - \frac{1}{\sqrt{2}}\omega^{\mu}_{N}K^{-} - \frac{1}{\sqrt{2}}\rho^{\mu0}K^{-} - \rho^{\mu-}\bar{K}^{0} + \frac{1}{\sqrt{2}}K^{\star\mu-}\eta_{N} - K^{\star\mu-}\eta_{S} \right. \\ &+ \left.\frac{1}{\sqrt{2}}K^{\star\mu-}\pi^{0} + \bar{K}^{\star\mu0}\pi^{-}\right) + K^{0}_{1,A\mu}\left(\omega^{\mu}_{S}\bar{K}^{0}\right. - \frac{1}{\sqrt{2}}\omega^{\mu}_{N}\bar{K}^{0} + \frac{1}{\sqrt{2}}\rho^{\mu0}\bar{K}^{0} - \rho^{\mu+}K^{-} + \frac{1}{\sqrt{2}}K^{\star\mu0}\eta_{N} \\ &- \bar{K}^{\star\mu0}\eta_{S} - \frac{1}{\sqrt{2}}K^{\star\mu0}\pi^{0} + \left.K^{\star\mu-}\pi^{+}\right) + K^{-}_{1,A\mu}\left(\frac{1}{\sqrt{2}}\omega^{\mu}_{N}K^{+} - \omega^{\mu}_{S}K^{+}\right. + \frac{1}{\sqrt{2}}\rho^{\mu0}K^{+} + \rho^{\mu+}K^{0} \\ &- \frac{1}{\sqrt{2}}K^{\star\mu+}\eta_{N} + K^{\star\mu+}\eta_{S} - \left.\frac{1}{\sqrt{2}}K^{\star\mu+}\pi^{0} - K^{\star\mu0}\pi^{+}\right) + \bar{K}^{0}_{1,A\mu}\left(\frac{1}{\sqrt{2}}\omega^{\mu}_{N}K^{0} - \omega^{\mu}_{S}K^{0}\right. - \frac{1}{\sqrt{2}}\rho^{\mu0}K^{0} \\ &+ \rho^{\mu-}K^{+} - \frac{1}{\sqrt{2}}K^{\star\mu0}\eta_{N} + K^{\star\mu0}\eta_{S} + \left.\left.\frac{1}{\sqrt{2}}K^{\star\mu0}\pi^{0} - K^{\star\mu+}\pi^{-}\right)\right\} \end{aligned} \label{lagalong}%
\end{equation}
and the explicit expression for the pseudovector part is given by
\begin{equation}
\begin{aligned} \mathcal{L}_{B} = \; &b\;\mathrm{Tr}\left\{ B_{\mu}\left[ V^{\mu},P\right]_{+}\right\} \\ = \; &\frac{b}{2\sqrt{2}} \left\{\frac{1}{\sqrt{2}}\right. f_{1N,B}\left(2\omega^{\mu}_{N}\eta_{N} + 2\rho^{\mu0}\pi^{0} + 2\rho^{\mu+}\pi^{-} + 2\rho^{\mu-}\pi^{+} + K^{\star\mu+}K^{-} + K^{\star\mu-}K^{+} + K^{\star\mu0}\bar{K}^{0} \right.\\ &+ \left.\bar{K}^{\star\mu0}K^{0}\right) + f_{1S,B}\left(2\omega^{\mu}_{S}\eta_{S} + K^{\star\mu-}K^{+} + K^{\star\mu+}K^{-} + K^{\star\mu0}\bar{K}^{0} + \bar{K}^{\star\mu0}K^{0}\right) + \frac{1}{\sqrt{2}}b^{0}_{1\mu}\left(2\omega^{\mu}_{N}\pi^{0} \right. \\ &+ \left.2\rho^{\mu0}\eta_{N} + K^{\star\mu+}K^{-} + K^{\star\mu-}K^{+} - K^{\star\mu0}\bar{K}^{0} - \bar{K}^{\star\mu0}K^{0}\right) + b^{+}_{1\mu}\left(\right.\sqrt{2}\rho^{\mu-}\eta_{N} + \sqrt{2}\omega_{N}\pi^{-} \\ &+ K^{\star\mu0}K^{-} + \left.K^{\star\mu-}K^{0}\right) + b^{-}_{1\mu}\left(\sqrt{2}\rho^{\mu+}\eta_{N} + \sqrt{2}\omega_{N}\pi^{+} + K^{\star\mu+}\bar{K}^{0} + \bar{K}^{\star\mu0}K^{+}\right) + K^{+}_{1,B\mu}\left(\omega^{\mu}_{S}K^{-}\right. \\ &+ \frac{1}{\sqrt{2}}\omega^{\mu}_{N}K^{-} + \frac{1}{\sqrt{2}}\rho^{\mu0}K^{-} + \rho^{\mu-}\bar{K}^{0} + \frac{1}{\sqrt{2}}K^{\star\mu-}\eta_{N} + K^{\star\mu-}\eta_{S} + \frac{1}{\sqrt{2}}K^{\star\mu-}\pi^{0} + \left.\bar{K}^{\star\mu0}\pi^{-}\right)\\ &+ K^{0}_{1,B\mu}\left(\omega^{\mu}_{S}\bar{K}^{0}\right. + \frac{1}{\sqrt{2}}\omega^{\mu}_{N}\bar{K}^{0} - \frac{1}{\sqrt{2}}\rho^{\mu0}\bar{K}^{0} + \rho^{\mu+}K^{-} + \frac{1}{\sqrt{2}}\bar{K}^{\star\mu0}\eta_{N} + \bar{K}^{\star\mu0}\eta_{S} - \frac{1}{\sqrt{2}}\bar{K}^{\star\mu0}\pi^{0} \\ &+ \left.K^{\star\mu-}\pi^{+}\right) + K^{-}_{1,B\mu}\left(\frac{1}{\sqrt{2}}\omega^{\mu}_{N}K^{+} + \omega^{\mu}_{S}K^{+}\right. + \frac{1}{\sqrt{2}}\rho^{\mu0}K^{+} + \rho^{\mu+}K^{0} + \frac{1}{\sqrt{2}}K^{\star\mu+}\eta_{N} + K^{\star\mu+}\eta_{S} \\ &+ \left.\frac{1}{\sqrt{2}}K^{\star\mu+}\pi^{0} + K^{\star\mu0}\pi^{+}\right) + \bar{K}^{0}_{1,B\mu}\left(\frac{1}{\sqrt{2}}\omega^{\mu}_{N}K^{0} + \omega^{\mu}_{S}K^{0}\right. - \frac{1}{\sqrt{2}}\rho^{\mu0}K^{0} + \rho^{\mu-}K^{+} \\ &+ \frac{1}{\sqrt{2}}K^{\star\mu0}\eta_{N} + K^{\star\mu0}\eta_{S} - \left.\left.\frac{1}{\sqrt{2}}K^{\star\mu0}\pi^{0} - K^{\star\mu+}\pi^{-}\right)\right\} \text{ .} \end{aligned} \label{lagblong}%
\end{equation}
The explicit form of the mixing part $\mathcal{L}_{mix}$ in the isospin
symmetric limit is given by Eq. (\ref{mixlag2}).


\bigskip

\begin{thebibliography}{99}                                                                                               %




\bibitem {amslerrev}C.~Amsler and N.~A.~Tornqvist,
Phys.\ Rept.\ \textbf{389}, 61 (2004).
E.~Klempt and A.~Zaitsev,
Phys. Rept.\ \textbf{454} (2007) 1


\bibitem {klempt}
E.~Klempt and A.~Zaitsev,
Phys.\ Rept.\ \textbf{454} (2007) 1 [arXiv:0708.4016 [hep-ph]].


\bibitem {dynrec}F.~Giacosa,
Phys.\ Rev.\ \textbf{D80 } (2009) 074028. [arXiv:0903.4481 [hep-ph]].
F.~Giacosa,
AIP Conf.\ Proc.\ \textbf{1322 } (2010) 223-231. [arXiv:1010.1021 [hep-ph]].


\bibitem {rupperice}
G.~Rupp, S.~Coito and E.~van Beveren,
Prog.\ Part.\ Nucl.\ Phys.\ \textbf{67} (2012) 449 [arXiv:1111.6850
[hep-ph]].


\bibitem {burakovsky}
L.~Burakovsky and J.~T.~Goldman,
Phys.\ Rev.\ D \textbf{57} (1998) 2879 [hep-ph/9703271].


\bibitem {scalars}
C.~Amsler and F.~E.~Close,
Phys.\ Rev.\ D \textbf{53} (1996) 295 [arXiv:hep-ph/9507326].
W.~J.~Lee and D.~Weingarten,
Phys.\ Rev.\ D \textbf{61}, 014015 (2000). [arXiv:hep-lat/9910008];
F.~E.~Close and A.~Kirk,
Eur.\ Phys.\ J.\ C \textbf{21}, 531 (2001). [arXiv:hep-ph/0103173].
F.~Giacosa, T.~Gutsche, V.~E.~Lyubovitskij and A.~Faessler,
Phys.\ Rev.\ D \textbf{72}, 094006 (2005). [arXiv:hep-ph/0509247].
F.~Giacosa, T.~Gutsche and A.~Faessler,
Phys. Rev. C \textbf{71}, 025202 (2005) [arXiv:hep-ph/0408085].
H.~Y.~Cheng, C.~K.~Chua and K.~F.~Liu,
Phys.\ Rev.\ D \textbf{74} (2006) 094005 [arXiv:hep-ph/0607206].
V.~Mathieu, N.~Kochelev and V.~Vento,
Int.\ J.\ Mod.\ Phys.\ E \textbf{18} (2009) 1 [arXiv:0810.4453 [hep-ph]].
F.~Giacosa,
Phys.\ Rev.\ D \textbf{75} (2007) 054007 [hep-ph/0611388].
A.~H.~Fariborz, R.~Jora and J.~Schechter,
Phys.\ Rev.\ D \textbf{72} (2005) 034001.
A.~H.~Fariborz,
Int.\ J.\ Mod.\ Phys.\ A \textbf{19} (2004) 2095.
M.~Napsuciale and S.~Rodriguez,
Phys.\ Rev.\ D \textbf{70} (2004) 094043.


\bibitem {otherscalars}E.~van Beveren, T.~A.~Rijken, K.~Metzger, C.~Dullemond,
G.~Rupp and J.~E.~Ribeiro,
Z.\ Phys.\ C \textbf{30} (1986) 615 [arXiv:0710.4067 [hep-ph]].
N.~A.~Tornqvist,
Z.\ Phys.\ C \textbf{68} (1995) 647 [hep-ph/9504372].
M.~Boglione and M.~R.~Pennington,
Phys.\ Rev.\ D \textbf{65} (2002) 114010 [hep-ph/0203149].
E.~van Beveren, D.~V.~Bugg, F.~Kleefeld and G.~Rupp,
Phys.\ Lett.\ B \textbf{641} (2006) 265 [hep-ph/0606022].
J.~R.~Pelaez,
Phys.\ Rev.\ Lett.\ \textbf{92} (2004) 102001.



\bibitem {denis}D.~Parganlija, F.~Giacosa, D.~H.~Rischke,
Phys.\ Rev.\ \textbf{D82 } (2010) 054024. [arXiv:1003.4934 [hep-ph]].
S.~Janowski, D.~Parganlija, F.~Giacosa, D.~H.~Rischke,
Phys.\ Rev.\ \textbf{D84 } (2011) 054007. [arXiv:1103.3238 [hep-ph]].
S.~Gallas, F.~Giacosa, D.~H.~Rischke,
Phys.\ Rev.\ \textbf{D82 } (2010) 014004. [arXiv:0907.5084 [hep-ph]].

\bibitem {dick}
D.~Parganlija, P.~Kovacs, G.~Wolf, F.~Giacosa and D.~H.~Rischke,
Phys.\ Rev.\ D \textbf{87} (2013) 014011 [arXiv:1208.0585 [hep-ph]].


\bibitem {latticescalar}
Y.~Chen, A.~Alexandru, S.~J.~Dong, T.~Draper, I.~Horvath, F.~X.~Lee, K.~F.~Liu
and N.~Mathur \textit{et al.},
Phys.\ Rev.\ D \textbf{73} (2006) 014516 [hep-lat/0510074].
S.~Prelovsek, T.~Draper, C.~B.~Lang, M.~Limmer, K.~-F.~Liu, N.~Mathur and
D.~Mohler,
Phys.\ Rev.\ D \textbf{82} (2010) 094507 [arXiv:1005.0948 [hep-lat]].
M.~Loan, X.~Q.~Luo and Z.~H.~Luo,
Int.\ J.\ Mod.\ Phys.\ A \textbf{21}, 2905 (2006) [arXiv:hep-lat/0503038];
E.~B.~Gregory, A.~C.~Irving, C.~C.~McNeile, S.~Miller and Z.~Sroczynski,
PoS \textbf{LAT2005}, 027 (2006) [arXiv:hep-lat/0510066];
Y.~Chen \textit{et al.},
Phys.\ Rev.\ D \textbf{73}, 014516 (2006) [arXiv:hep-lat/0510074];
E.~Gregory, A.~Irving, B.~Lucini, C.~McNeile, A.~Rago, C.~Richards and
E.~Rinaldi,
JHEP \textbf{1210} (2012) 170 [arXiv:1208.1858 [hep-lat]].
M.~Wagner, C.~Alexandrou, J.~O.~Daldrop, M.~D.~Brida, M.~Gravina, L.~Scorzato,
C.~Urbach and C.~Wiese,
arXiv:1302.3389 [hep-lat].

\bibitem {quarkmodel}S.~Godfrey and N.~Isgur,
Phys.\ Rev.\ D \textbf{32} (1985) 189.
See also the note on the quark model in Ref. \cite{PDG}.

\bibitem {tensor}
F.~Giacosa, T.~Gutsche, V.~E.~Lyubovitskij, A.~Faessler,
Phys.\ Rev.\ D \textbf{72} (2005) 114021 [hep-ph/0511171].


\bibitem {tensorlast}
Z.~-C.~Ye, X.~Wang, X.~Liu and Q.~Zhao,
Phys.\ Rev.\ D \textbf{86} (2012) 054025 [arXiv:1206.0097 [hep-ph]].

\bibitem {coitoacta}
G.~Rupp, S.~Coito and E.~van Beveren,
Acta Phys.\ Polon.\ Supp.\ \textbf{5} (2012) 1007 [arXiv:1209.1475 [hep-ph]].

\bibitem {geffen}S.~Gasiorowicz and D.~A.~Geffen,
Rev.\ Mod.\ Phys.\ \textbf{41}, 531 (1969).


\bibitem {ko}P.~Ko and S.~Rudaz,
Phys.\ Rev.\ D \textbf{50}, 6877 (1994). M.~Urban, M.~Buballa, J.~Wambach,
Nucl.\ Phys.\ \textbf{A697 } (2002) 338-371. [hep-ph/0102260].

\bibitem {dyngena1}M.~F.~M.~Lutz and E.~E.~Kolomeitsev,
Nucl.\ Phys.\ A \textbf{730} (2004) 392 [arXiv:nucl-th/0307039].
M.~Wagner and S.~Leupold,
Phys.\ Lett.\ B \textbf{670} (2008) 22 [arXiv:0708.2223 [hep-ph]].
M.~Wagner and S.~Leupold,
Phys.\ Rev.\ D \textbf{78} (2008) 053001 [arXiv:0801.0814 [hep-ph]].
S.~Leupold and M.~Wagner,
arXiv:0807.2389 [nucl-th].
L.~S.~Geng, E.~Oset, J.~R.~Pelaez and L.~Roca,
Eur.\ Phys.\ J.\ A \textbf{39} (2009) 81 [arXiv:0811.1941 [hep-ph]].


\bibitem {sukuki}
M.~Suzuki,
Phys.\ Rev.\ D \textbf{47} (1993) 1252.


\bibitem {cheng}
H.~-Y.~Cheng,
Phys.\ Rev.\ D \textbf{67} (2003) 094007 [hep-ph/0301198].


\bibitem {hatanaka}
H.~Hatanaka and K.~-C.~Yang,
Phys.\ Rev.\ D \textbf{77} (2008) 094023 [Erratum-ibid.\ D \textbf{78} (2008)
059902] [arXiv:0804.3198 [hep-ph]].


\bibitem {chengyang}
H.~-Y.~Cheng and K.~-C.~Yang,
Phys.\ Rev.\ D \textbf{76} (2007) 114020 [arXiv:0709.0137 [hep-ph]].


\bibitem {chengrevisiting}
H.~-Y.~Cheng,
Phys.\ Lett.\ B \textbf{707} (2012) 116 [arXiv:1110.2249 [hep-ph]].


\bibitem {hatanakanew}
H.~Hatanaka and K.~-C.~Yang,
Phys.\ Rev.\ D \textbf{78} (2008) 074007 [arXiv:0808.3731 [hep-ph]].


\bibitem {ahmednew}
A.~Ahmed, I.~Ahmed, M.~Ali Paracha and A.~Rehman,
Phys.\ Rev.\ D \textbf{84} (2011) 033010 [arXiv:1105.3887 [hep-ph]].


\bibitem {lihuanew}
Y.~Li, J.~Hua and K.~-C.~Yang,
Eur.\ Phys.\ J.\ C \textbf{71} (2011) 1775 [arXiv:1107.0630 [hep-ph]].


\bibitem {salam}P.~T.~Matthews and A.~Salam,
Phys.\ Rev.\ \textbf{112} (1958) 283.
P.~T.~Matthews and A.~Salam,
Phys.\ Rev.\ \textbf{115} (1959) 1079.


\bibitem {achasov}N.~N.~Achasov and A.~V.~Kiselev,
Phys.\ Rev.\ D \textbf{70} (2004) 111901 [arXiv:hep-ph/0405128].


\bibitem {lupo}F.~Giacosa and G.~Pagliara,
Phys.\ Rev.\ C \textbf{76} (2007) 065204 [arXiv:0707.3594 [hep-ph]].

\bibitem {duecan}F.~Giacosa,
Found.\ Phys.\ \textbf{42} (2012) 1262 [arXiv:1110.5923 [nucl-th]].
F.~Giacosa and T.~Wolkanowski,
Mod.\ Phys.\ Lett.\ A \textbf{27} (2012) 1250229 [arXiv:1209.2332 [hep-ph]].
F.~Giacosa and G.~Pagliara,
Nucl.\ Phys.\ A \textbf{812} (2008) 125 [arXiv:0804.1572 [hep-ph]].


\bibitem {kloe}
F.~Ambrosino, A.~Antonelli, M.~Antonelli, F.~Archilli, P.~Beltrame,
G.~Bencivenni, S.~Bertolucci and C.~Bini \textit{et al.},
JHEP \textbf{0907} (2009) 105 [arXiv:0906.3819 [hep-ph]].
G.~Amelino-Camelia, F.~Archilli, D.~Babusci, D.~Badoni, G.~Bencivenni,
J.~Bernabeu, R.~A.~Bertlmann and D.~R.~Boito \textit{et al.},
Eur.\ Phys.\ J.\ C \textbf{68} (2010) 619 [arXiv:1003.3868 [hep-ex]].

\bibitem {hanhart}
C.~Hanhart, B.~Kubis and J.~R.~Pelaez,
Phys.\ Rev.\ D \textbf{76} (2007) 074028 [arXiv:0707.0262 [hep-ph]].


\bibitem {k0}
R.~Battiston, D.~Cocolicchio, G.~L.~Fogli and N.~Paver,
Phys.\ Rept.\ \textbf{214} (1992) 293.


\bibitem {rupp}
S.~Coito, G.~Rupp and E.~van Beveren,
Phys.\ Rev.\ D \textbf{84} (2011) 094020 [arXiv:1106.2760 [hep-ph]].


\bibitem {PDG}J.~Beringer \textit{et al.} [Particle Data Group
Collaboration],
Phys.\ Rev.\ D \textbf{86} (2012) 010001.


\bibitem {wolkanowski}T.~Wolkanowski, `\textit{Resonances and poles in the
second Riemann sheet}', master thesis at the J. W. Goethe University,
Frankfurt am Main (Germany) [arXiv:1303.4657 [hep-ph]].

\bibitem {alekseev}M. G. Alekseev \textit{et al}, Phys. Rev. Lett.
\textbf{104} 241803 (2010).

\bibitem {atkinson}M. Atkinson \textit{et al., }Nucl. Phys. B \textbf{242} 269.

\bibitem {carnegie}R.K. Carnegie \textit{et al.}, Nucl. Phys. B \textbf{127} 509.
\end{thebibliography}
\end{document}